\documentclass[%
reprint,
showpacs,
amsmath,amssymb,
aps,
prb,
]{revtex4-1}

\usepackage{graphicx}
\usepackage{dcolumn}
\usepackage{bm}
\usepackage{graphicx}
\usepackage{dcolumn}
\usepackage{bm}
\usepackage{bigints}
\usepackage{color}
\usepackage{amsmath,amssymb,graphicx}
\usepackage{float}
\usepackage{braket}
\usepackage[utf8]{inputenc}
\usepackage[T1]{fontenc}
\usepackage{textcomp}
\usepackage{gensymb}

\begin{document}


\title{Random lasing in an Anderson localizing optical fiber}

\author{Behnam Abaie$^{1,2}$, Esmaeil Mobini$^{1,2}$, Salman Karbasi$^3$, Thomas Hawkins$^4$, John Ballato$^4$, and Arash~Mafi$^{1,2*}$}

\address{$^1$Department of Physics \& Astronomy, University of New Mexico, Albuquerque, NM 87131, USA.
\\
$^2$Center for High Technology Materials, University of New Mexico, Albuquerque, NM 87106, USA.
\\
$^3$Department of Electrical Engineering, University of California, San Diego, CA 92093, USA.
\\
$^4$Center for Optical Materials Science and Engineering Technologies (COMSET) and the Department of Materials Science and Engineering, Clemson University, Clemson, SC 29634, USA.
$^*$Corresponding author: mafi@unm.edu}

\date{\today}

\begin{abstract}
A directional random laser mediated by transverse Anderson localization in a disordered glass optical fiber is reported. Previous demonstrations of random lasers have found limited applications because of their multi-directionality and chaotic fluctuations in the laser emission. The random laser presented in this paper operates in the Anderson localization regime. The disorder induced localized states form isolated local channels that make the output laser beam highly directional and stabilize its spectrum. The strong transverse disorder and longitudinal invariance result in isolated lasing modes with negligible interaction with their surroundings, traveling back and forth in a Fabry-Perot cavity formed by the air-fiber interfaces. It is shown that if a localized input pump is scanned across the disordered fiber input facet, the output laser signal follows the transverse position of the pump. Moreover, a uniformly distributed pump across the input facet of the disordered fiber generates a laser signal with very low spatial coherence that can be of practical importance in many optical platforms including image transport with fiber bundles.
\end{abstract}
\maketitle
\section{Introduction}
Unlike conventional lasers that require a resonator cavity to operate, random lasers exploit multiple
scattering to trap light and provide feedback to the system~\cite{cao2005random, wiersma2008physics, Graydon}. 
One of the first observations of random lasing was emissions from a laser dye solution containing 
micro-particles~\cite{lawandy1994laser, Wiersma1995Nature,Lawandy1995Nature}. Since then, there have been 
several reports that attribute random lasing in disordered media either to diffusive extended modes or 
Anderson localized modes~\cite{wiersma2008physics, wiersma1996light, cao1999random, cao2000transition, cao2000spatial, 
cao2001photon, cao2003random, cao2003mode,lima2017observation}.
Anderson localization was first used by Pradhan and Kumar to show increased reflected intensity due to wave confinement in 
a mirrorless amplifying 1D structure~\cite{pradhan1994localization}. Later, enhancement of lasing in random multilayer stacks 
and planar waveguides was explained conceptually by Anderson localization~\cite{jiang2000time, vanneste2001selective}. 
The nature of lasing modes in disordered media, particularly the role of Anderson localization in these systems, 
is still a matter of debate~\cite{van2007spatial, tureci2008strong, wiersma2013disordered}. Currently, it is accepted that Anderson localization is not required for coherent random 
lasing in disordered media~\cite{fallert2009co, Fujii2011thesis}. However, Anderson localized lasing modes can result in a narrower frequency response analogous 
to closed cavities in regular lasers~\cite{stano2013suppression}. 

Here, we present a disordered laser system for which Anderson localization plays an integral role in determining its lasing characteristics--strong transverse wave confinement hosted by a dye-filled glass Anderson localizing optical fiber (g-ALOF) reduces spatial overlap of the lasing modes and results in high spectral stability of the random laser emission. The disordered glass fiber has a randomly distributed air-hole pattern in transverse dimensions which remains invariant along the length of the fiber for the 
typical lengths used in Ref.~\cite{karbasi2012transverse} and here. The air fill fraction was shown to be higher near the outer boundaries and therefore localization is stronger near the boundaries in comparison to the central regions of the fiber. For more details about the disorder structure of g-ALOF please see Supplementary Fig.~S4.

\begin{figure}[h]
  \centering
  \includegraphics[width=.85\linewidth]{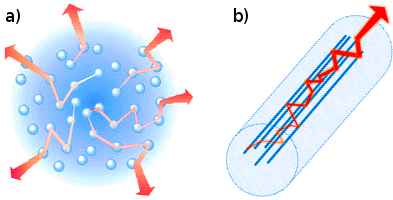}
\caption{\label{fig:Fig1}
{\small \bf  Directional random lasing in a g-ALOF.} \small a) Light from a random laser is usually multi-directional. b) Schematic of directional random lasing mediated by a disordered optical fiber.}
\end{figure}

The output of a random laser is usually multi-directional which can hinder the usefulness of such sources (Fig.~\ref{fig:Fig1}a). In one of the first efforts to tackle this problem, active control of the spatial distribution of the pump beam was proposed to achieve a highly directional random laser~\cite{hisch2013pump}. More recently, Sch{\"o}nhuber et al.~\cite{schonhuber2016random,wiersmaNature2016} designed a collimated random laser beam by sandwiching an active region between a bottom metallic layer and a planar waveguide patterned with random air-holes to extract the light. Our results present an alternative robust design where transversely Anderson localized (TAL) modes of a dye-filled disordered optical fiber are exploited to obtain directional random lasing; transverse light confinement due to disorder-induced localized modes combined with free propagation in the longitudinal direction result in a highly directional random laser 
output (Fig.~\ref{fig:Fig1}b).
\section{Materials and methods}
\subsection{Fabrication of the fiber}
g-ALOF used in this work is drawn from “satin quartz” (Heraeus Quartz), which is a porous artisan glass. The initial rod is 8~mm in diameter and 850~mm in length and is drawn at Clemson University on a Heathway draw tower at a temperature of 1890$\degree C$. The tip of the fiber is imaged using a Hitachi SU-6600 analytical variable pressure field emission scanning electron microscope (SEM). The diameter of the fiber is approximately 250~\textmu m and the average air fill-fraction is about 5.5\% with the air-hole diameters varying between about 0.2~\textmu m to 5.5~\textmu m. 
\subsection{Active fiber preparation and measurements}
The outer acrylate coating of g-ALOF is removed using a hot air gun. A piece of g-ALOF with approximately 10~cm length is cleaved by a York PK Technology fiber cleaver and dipped inside a rhodmaine 640 solution in ethanol or benzyl alcohol (or a mixture of both) at a concentration of 0.5~mg/ml for about 6 hours. The tip of the dye-filled g-ALOF is investigated under a microscope to ensure the holes are uniformly filled with the dye solution. A 10~mm piece of the dye-filled g-ALOF is cleaved and mounted on an adjustable fiber clamp to be used in the experiments. The dye-filled g-ALOF is end-pumped by a frequency doubled Nd:YAG laser with a pulse duration of 0.6~ns and a repetition rate of 50~Hz, using a 20X microscope objective (OBJ1 in Supplementary Fig. S5). The repetition rate and pump energy is kept low to delay the optical bleaching of the dye. The near-field image at the output tip of active g-ALOF is imaged on a CCD beam profiler by a 40X microscope objective (OBJ2 in Supplementary Fig. S5). 
\subsection{Numerical simulations}
The guided modes of g-ALOF are calculated using COMSOL Multiphysics~\cite{multiphysics2016version}. The Floating Network Licensed (FNL) software is installed on a cluster located at the Center for Advanced Research Computing of the University of New Mexico, meeting large memory requirements for the heavy calculations. The geometry of g-ALOF is imported to COMSOL using the SEM image of the tip of the fiber by rendering the image into a compatible vector format (DXF) using Inkscape Vector Graphics Editor software. The refractive index profile of g-ALOF, used in the simulations, is shown in Supplementary Fig. S6.
\subsection{Characterization of the spectral stability}
In order to characterize the spectral stability of the laser in the Anderson localized regime, the spectrum is measured in a sequential mode such that 100 data acquisitions are done in 3 seconds (each spectrum is integrated over 30~ms). Such a fast data acquisition provides the possibility to compare the laser emission spectra under excitations with individual shots of a pump pulse train with 50~Hz repetition rate. In order to quantify the fluctuations of the laser spectrum under excitations with individual pump pulses, we use the Normalized Mean Integrated Squared Error (NMISE) defined by
\begin{align}
\label{eq:NMISE}
	NMISE = \dfrac{\sum\limits_{i=1}^N \int(f_i(\lambda)-\bar{f}(\lambda))^2d\lambda}{N\times \int\bar{f}(\lambda)^2d\lambda},
\end{align}
where $f_1(\lambda), f_2(\lambda),…, f_N(\lambda)$ are the data collected for the laser spectrum at $N$ successive identical pump pulses, and $\bar{f}(\lambda)$ is the average of them. NMISE ranges between 0 and 100\%, where 0\% is proportional to a series of completely identical spectra and 100\% is related to a series of completely distinct spectra.
\section{Results and discussion}
\subsection{Transversely localized lasing}
\begin{figure}[h]
\centerline{
\includegraphics[width=8cm]{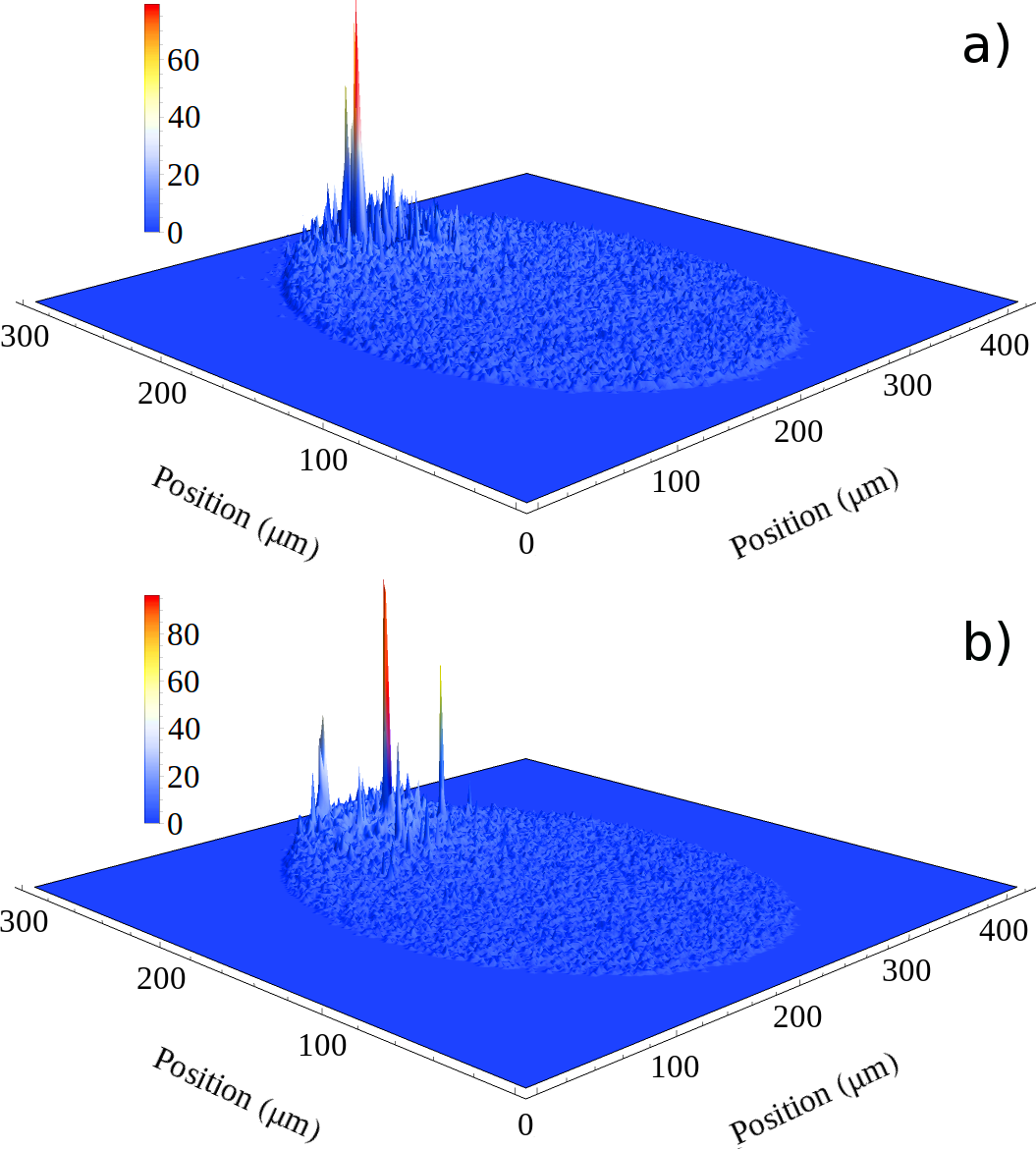}
}
\caption{\label{fig:BP1}
{\small \bf  TAL of pump and laser in the mixed regime.} \small Anderson localization of the: a) pump beam near the edge, and b) laser beam associated with the pump condition in part (a). Narrow peaks in the laser beam profile is due to enhancement of lasing mediated by step-index guiding inside the individual holes of the active g-ALOF.}
\end{figure}

The pump and laser beam profiles at the output facet of g-ALOF are shown in Fig.~\ref{fig:BP1}. The pump beam at the input facet is coupled near the edge of g-ALOF. The residual pump in the output remains clamped in the same transverse location, consistent with Anderson localization of g-ALOF presented in Ref.~\cite{karbasi2012transverse}. The laser beam profile, shown in Fig.~\ref{fig:BP1}b, is also localized at the same transverse position. Note that in these results, the dye (benzyl solution) has a higher refractive index than the host glass; therefore, step-index guiding is playing a role for individual dye filled air-holes of g-ALOF. The stimulated emission is enhanced in these regions due to a stronger interaction of laser beam and gain; therefore, step-index guiding manifests itself in the form of strong discrete peaks in the laser beam profile in Fig.~\ref{fig:BP1}b. Nevertheless, TAL is present at least in the form of a localized pump beam. Because the hallmarks of both TAL and step-index guiding are simultaneously present in these results, this regime of laser operation is referred to as the “mixed regime”. In the following, results that are solely due to TAL are presented for which the step-index guiding is completely absent and the spatial behavior of the laser beam is entirely dictated by Anderson localization.

\begin{figure}[h]
\centerline{
\includegraphics[width=8cm]{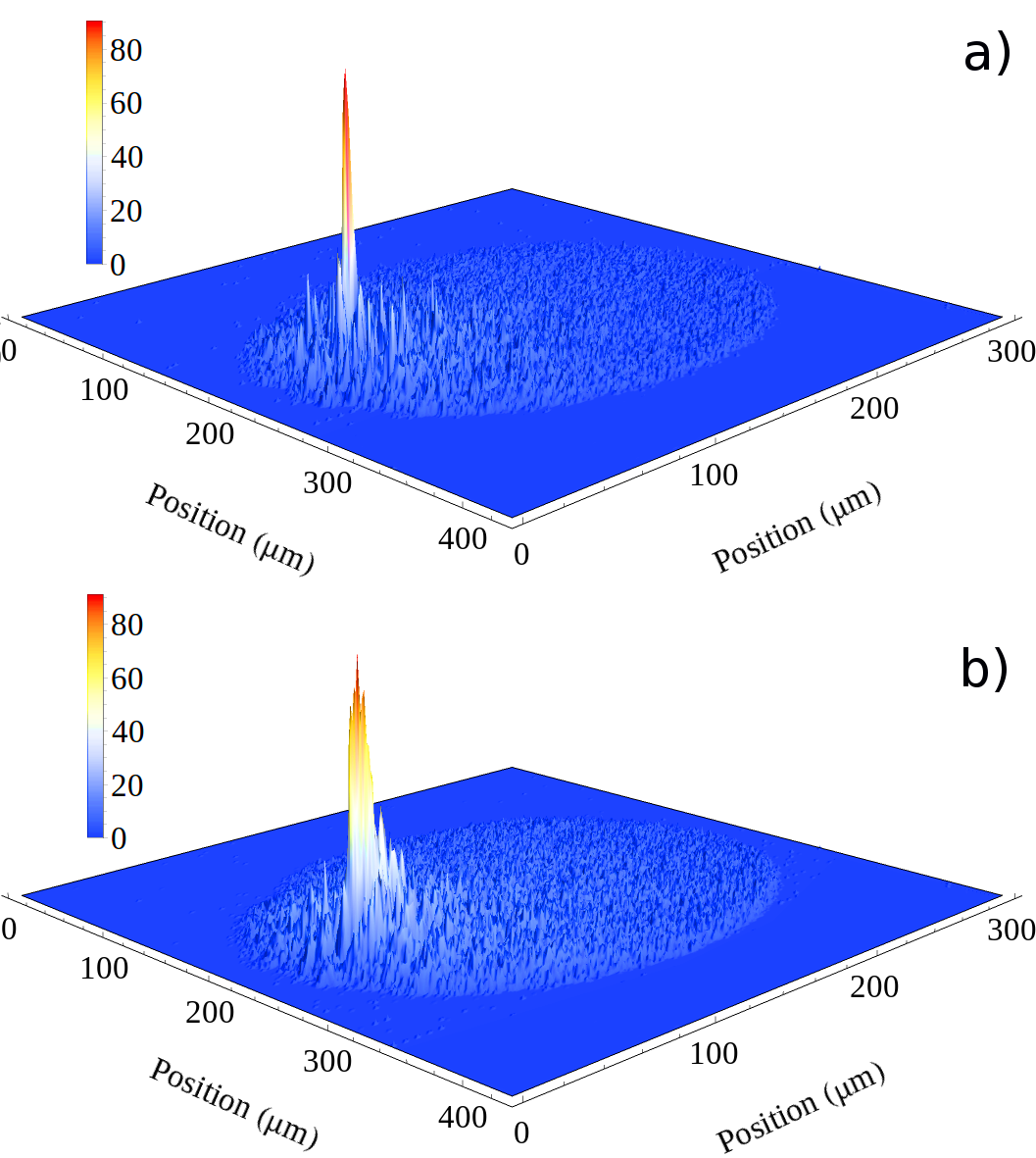}
}
\caption{\label{fig:BP2}
{\small \bf TAL of pump and laser in the Anderson localized regime.} \small Anderson localization of the: a) pump beam near the edge, and b) laser beam associated with the pump condition in part (a). The air-holes of g-ALOF are filled with rhodmaine 640 solution in ethanol; therefore, step-index guiding is not playing a role. Laser beam profile is transversely localized and follows the transverse position of the pump.}
\end{figure}

In pursuance of evading step-index guiding, the air-holes are filled with rhodmaine 640 solution in ethanol. The refractive index of ethanol (1.37) is smaller than that of the host glass (1.46); therefore, individual air holes filled with the solution do not form local waveguides in the active medium. Figure~\ref{fig:BP2} shows the transverse localization of the pump and laser near the edge of g-ALOF filled with this solution. This behavior is fully dictated by the TAL of the pump and laser: in the absence of TAL, the focused pump beam coupled near the edge of g-ALOF would have quickly diffracted and covered the entire cross section of g-ALOF, given that the Rayleigh range $Z_R$ for the focused pump is on the order of $Z_R\sim$ 200~\textmu m. Figure~\ref{fig:BP1}a shows that TAL clearly dominates diffraction mediated by the strong transverse disorder in g-ALOF and confines the pump beam around its incoming transverse position as it freely propagates along the fiber. Notice that, based on Ref.~\cite{karbasi2012transverse}, the effective beam radius of the pump expands as the beam propagates along the fiber until it reaches its final localized value, after which it does not change substantially. Numerically it was shown that the stabilized effective beam radius is reached after about 30 mm of propagation in g-ALOF. 

The active g-ALOF used here is 10~mm long and the scattering strength is also reduced due to the dye filling, effectively leading to a longer stabilization distance. Therefore, the effective beam radius of the residual pump at the output facet of g-ALOF is not fully stabilized. Nevertheless, it is quite close to the stabilized value~\cite{karbasi2012transverse}. The laser beam profile associated with the pump condition described above is shown in Fig.~\ref{fig:BP2}b. It is transversely localized and highly resembles the pump beam profile. However, the laser effective beam radius is at its stabilized form because of the laser stabilization condition which requires multiple round trips and amplification in the cavity. Notice that the localized laser beam in an active g-ALOF requires a localized pump beam, because g-ALOF supports numerous localized modes that are distributed across the transverse dimensions of the fiber and an extended non-localized pumping can excite many of these localized modes simultaneously, resulting in an extended output laser signal. We emphasize that in a non-disordered fiber without mode localization, even localized pumping does not result in localized lasing.
In Figs.~\ref{fig:BP1} and \ref{fig:BP2}, the output laser beam follows the transverse position of the input pump. Localized states in g-ALOF trap the beam, which is propagating back and forth between air-fiber interfaces. In other words, the disorder induced localized states form several isolated channels located across the transverse dimension of g-ALOF. Upon excitation of one of these channels by a narrow input pump, the system starts lasing by the feedback provided through the 4\% reflection at each air-fiber interface. Therefore, the presence of the TAL and the Fabry-Perot formed by the air-fiber interfaces results in a directional random laser. Evaluation of g-ALOF laser beam quality based on the variance method~\cite{siegman1998wow,mafi2005beam} followed by a short discussion about the directionality is provided in Supplementary. 

\begin{figure*}[t]
\centerline{
\includegraphics[width=16cm]{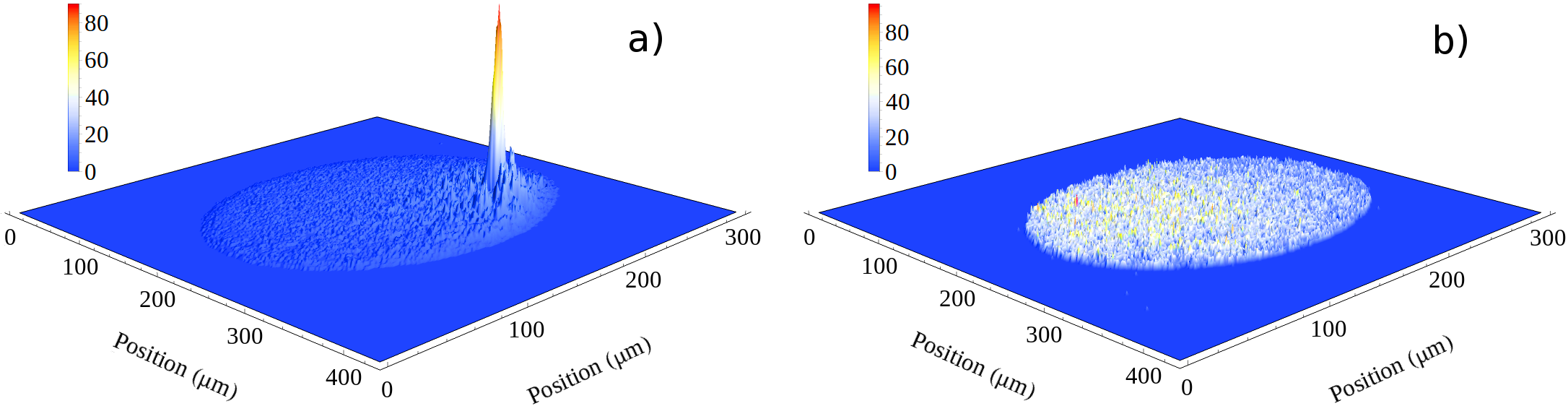}
}
\caption{\label{fig:BP3}
{\small \bf Stronger localized lasing near the edge of g-ALOF.} \small Localization of the laser is stronger near the boundaries. A narrow pump beam is scanned across the input facet of g-ALOF to excite disorder induced channels a) near the edge, and b) at the center of the disordered fiber, respectively. Stronger TAL near the edge results in a narrower output laser in comparison to the central regions.}
\end{figure*}

As we mentioned earlier, TAL is stronger near the boundaries of g-ALOF in comparison to the central regions of the fiber due to the higher disorder in these regions~\cite{karbasi2012transverse}. In order to compare the results with previous analysis on passive g-ALOF, the pump beam is launched at two different transverse positions of the active g-ALOF by scanning the pump objective using a precision XYZ translation stage. Subfigures~\ref{fig:BP3}a and b, show the laser beam profile at the output facet of g-ALOF when the input pump beam is launched near the edge and the center of g-ALOF input facet, respectively (residual pump beam profiles are not exhibited in this Figure). Clearly, Anderson localization occurs more strongly around the edge of g-ALOF in agreement with the results reported in Ref.~\cite{karbasi2012transverse}. Because the presented results are solely due to TAL and step-index guiding is absent, we call this regime of laser operation as “Anderson-localized regime” in the sections that follow.
\section{Origin of localized lasing}
\subsection{Localized lasing modes}
In order to further investigate the presence of localized lasing modes in active g-ALOF, we have numerically calculated guided modes of the system using the Finite Element Method (FEM). A small imaginary part is added to the refractive index of the dye-filled air-holes to represent the gain. The refractive index profile used for simulations is shown in Supplementary Fig. S6. If calculated modes are Anderson localized, the system can potentially support localized lasing. Here, we only show results in the Anderson-localized regime of laser operation, but similar results have been verified for the mixed regime as well.

\begin{figure}[h]
\centerline{
\includegraphics[width=\columnwidth]{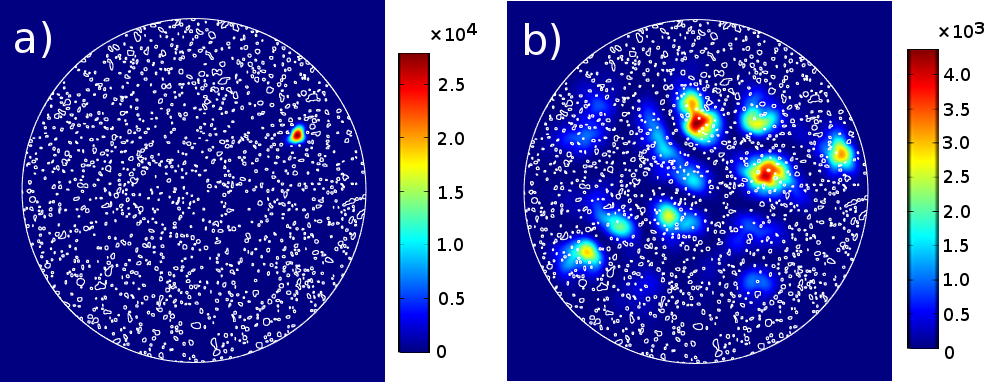}
}
\caption{\label{fig:Modes}
{\small \bf Calculated localized lasing modes.} \small Time average power of typical calculated lasing modes. a) The localization length is smaller than the system size due to strong transverse disorder and TAL. b) Reduced scattering strength, in comparison to part a, leads to a weaker TAL, and therefore localization length is larger than the system size and the mode nearly covers the entire transverse dimensions of g-ALOF.}
\end{figure}

The simulations are performed in the frequency domain for a wavelength that falls inside the emission spectrum of the laser (around 610 nm). The refractive index of the host glass is set at 1.46, and that of the dye-filled air-holes is set at 1.37 with a small imaginary part. Fig.~\ref{fig:Modes}a shows the time averaged power flow of a typical calculated mode. The background pattern is the transverse geometry of g-ALOF extracted directly from its SEM image. The calculated mode is clearly localized due to the presence of transverse disorder and strong multiple scattering. We note that other methods such as local step-index guiding or photonic bandgap guiding can also be used to localize light but are not playing a role in the structure studied here~\cite{mafi2015transverse,karbasi2012transverse}. The impact of TAL can be verified by decreasing the refractive index contrast between the dye-filled air-holes and the host glass. In this regard, Fig.~\ref{fig:Modes}b shows a typical guided mode with exactly the same configuration as Fig.~\ref{fig:Modes}a except with the refractive index of the dye-filled air-holes set at 1.459. Decreasing the refractive index contrast and therefore scattering strength has clearly reduced the impact of TAL such that the localization radius of the mode is larger than the system size, so it appear to be nearly extended over the entire transverse dimensions of g-ALOF. More examples of localized lasing modes are presented in Supplementary Fig. S7.
\subsection{Localization in passive g-ALOF}
TAL in g-ALOF has been rigorously explored in Ref.~\cite{karbasi2012transverse}. However, the reported results were for a passive g-ALOF without any dye filling. It is important to note that filling the air holes of g-ALOF with ethanol or benzyl alcohol reduces their refractive index contrast with the host glass which causes a weaker TAL~\cite{karbasi2012detailed}. Another important difference is in the wavelengths of the laser and pump used here in comparison to the 405~nm diode laser used in Ref.~\cite{karbasi2012transverse}. Here, we report TAL in g-ALOF filled with ethanol. Results are similar to the case of g-ALOF filled with benzyl alcohol, which are not reported here. We emphasize that the ethanol used here is pure ($\sim 99\%$ purity) with no dye solute. A He-Ne laser (center wavelength $\sim 633~nm$) is used to carry out the experiments, because its wavelength is in the emission range of g-ALOF laser that is reported later in this paper (see Fig.~\ref{fig:Stability}).

The experimental setup is similar to Supplementary Fig. S5 except that the Nd:YAG laser is replaced with the He-Ne laser, and g-ALOF used in the experiment is filled with pure ethanol. Figure~\ref{fig:PassiveBP}a shows near field image of the tip of g-ALOF at the output. The input objective (OBJ1 in Supplementary Fig. S5) is scanned across the input facet of g-ALOF to compare localization near the boundaries with the central regions. Clearly, localization is stronger when the objective launches the input beam near the boundary of g-ALOF, shown in Fig.~\ref{fig:PassiveBP}a, compared with Fig.~\ref{fig:PassiveBP}b, where the input beam is launched near the center. These results show that g-ALOF supports strongly localized modes in the spectral emission range of g-ALOF laser even when the air holes are filled with ethanol, in comparison with Ref.~\cite{karbasi2012transverse} where the air holes are open. 
\begin{figure*}[t]
\centerline{
\includegraphics[width=16cm]{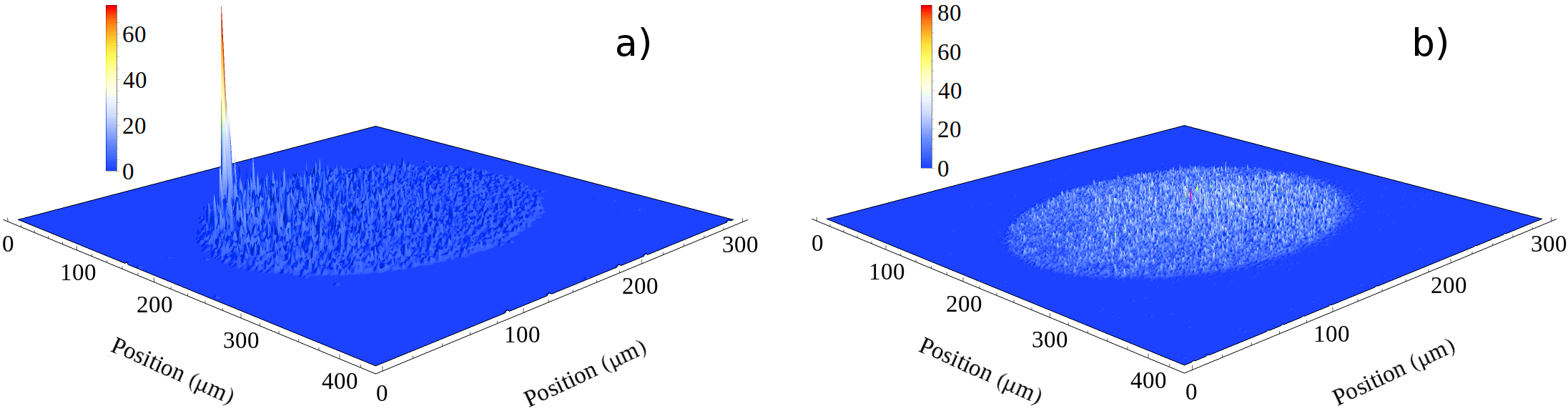}
}
\caption{\label{fig:PassiveBP}
{\small \bf Localization in passive g-ALOF.} \small TAL in passive g-ALOF filled with ethanol: a) near the edge, and b) at the center. The disordered fiber supports strongly localized modes in the spectral emission range of g-ALOF laser even when the air holes are filled with ethanol, in comparison with Ref.~\cite{karbasi2012transverse} where the air holes are open. TAL is stronger near the boundary of g-ALOF in agreement with Ref.~\cite{karbasi2012transverse}.}
\end{figure*}
\section{Laser spectrum}
\subsection{Anderson-localized regime}
In Ref.~\cite{mujumdar2007chaotic}, lasing in a dye filled porous glass disk was reported. The scattering strength in the system was weak and therefore the results were in the diffusive regime. It was shown that, the emission spectra of such an amplifying disordered medium were distinct and uncorrelated at each shot of the identical pump pulses. This behavior was explained based on the strong coupling of modes initiated by spontaneous emission, and Anderson localized regime was suggested to reduce mode competition and chaotic behavior of the random laser. Here, we study the stability of g-ALOF laser emission spectra. Our results show that, for a strongly localized lasing mode, the spectrum remains nearly unchanged at consecutive shots of the pump pulse train.

\begin{figure}[t]
\centerline{
\includegraphics[width=0.975\columnwidth]{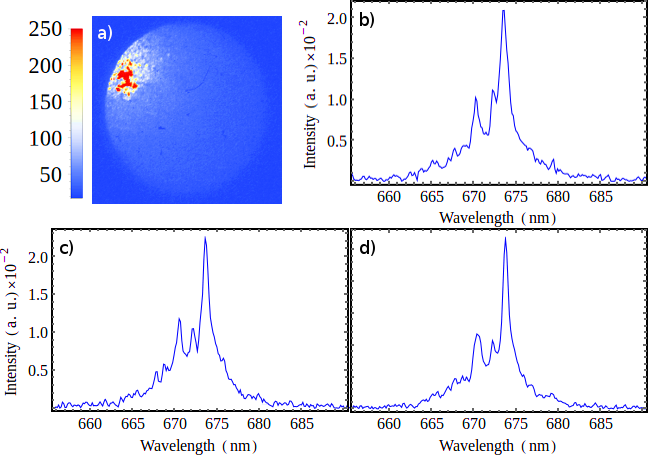}
}
\caption{\label{fig:Stability}
{\small \bf Laser emission spectrum in the Anderson localized regime.} \small a) a strongly localized lasing mode and its associated spectra under excitation with three individual pump pulses in b, c, and d. The spectrum shows high stability as the narrow spikes remain the same under excitation with individual pump pulses in contrast to the results presented in Ref.~\cite{mujumdar2007chaotic}.}
\end{figure}

Figure~\ref{fig:Stability} shows a strongly localized lasing mode and the associated emission spectra;  the emission spectrum under pumping with three successive identical pulses are shown in subfigures~\ref{fig:Stability}b, c, and d. The spectrum shows a relatively high stability; the narrow spikes remain at the same wavelength in contrast to Ref.~\cite{mujumdar2007chaotic} where the spikes appeared at distinct wavelengths under excitation with identical pump pulses. In order to quantify the stability of g-ALOF laser spectrum, we use NMISE (see Materials and Methods), as a measure of the spectrum fluctuations over 100 successive identical pump pulses. The calculated value of NMISE for the strongly localized mode in Fig.~\ref{fig:Stability}a is very small ($\sim$4\%) which indicates the high spectral stability. Further information about the dependency of the spectral stability to the localization strength of the lasing modes is presented in Supplementary.  

High spectral stability of g-ALOF laser is understood based on the strong mode confinement granted by the localized states; as discussed earlier, disorder induced localized states form guiding channels located across the transverse dimension of the disordered fiber. Once a narrow pump beam excites one of these channels, the system starts lasing by the feedback provided through reflections at air-fiber interfaces. TAL reduces mode competition which is the underlying mechanism that causes chaotic fluctuations in the emission spectra of random lasers with weak mode confinement such as the one reported in Ref.~\cite{mujumdar2007chaotic}.
\subsection{Mixed regime}
\begin{figure}[t]
\centerline{
\includegraphics[width=0.85\columnwidth]{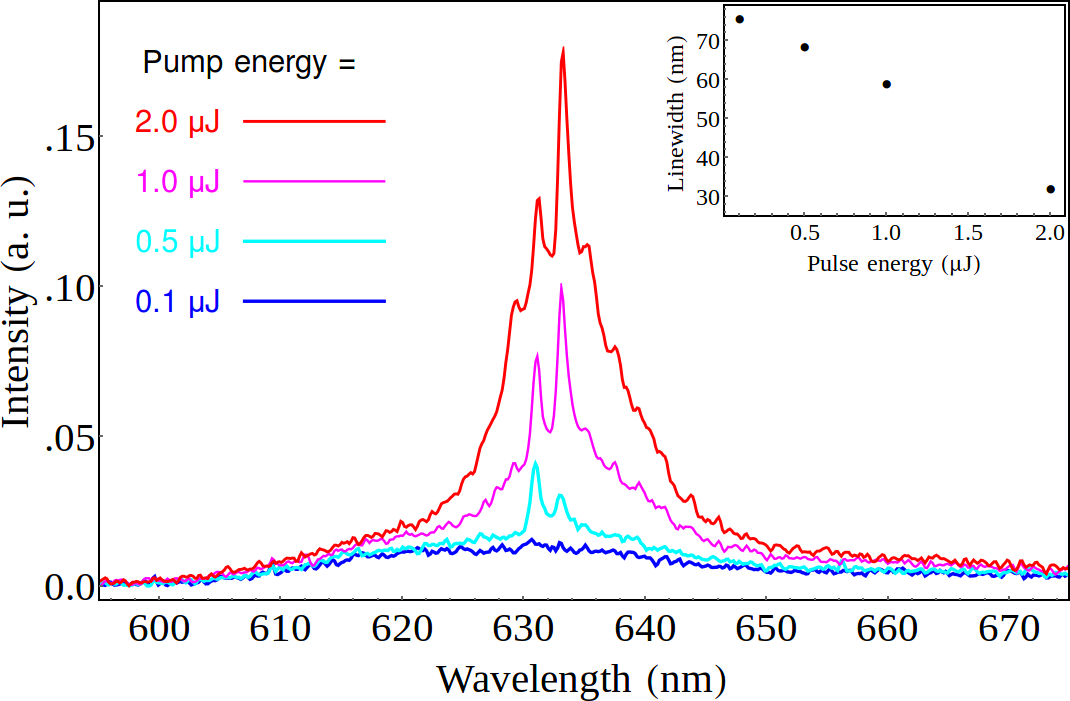}
}
\caption{\label{fig:MixedRegimeSpectrum}
{\small \bf Laser emission spectrum in the mixed regime.} \small Narrow spikes on top a global narrowing of the spectrum can be seen. Emission spectrum in this regime of laser operation is highly sensitive to the pump position due to the impact of step-index guiding inside local waveguides with diverse geometries across the disordered fiber. Inset shows the linewidth narrowing with respect to the pump energy.}
\end{figure}

Figure~\ref{fig:MixedRegimeSpectrum} shows the emission spectrum in the mixed regime at different pump pulse energy levels. Inset shows the linewidth narrowing with respect to the pump energy. As expected, the laser spectrum narrows as the pump power increases; narrow spikes on top of the global narrowing of the laser spectrum is evident. In this regime of laser operation, the emission spectrum depends highly on the pump condition, where scanning the pump across the fiber input facet affects the spectrum significantly (not shown here). This is due to the local waveguides formed at various transverse positions of the fiber as explained previously. Because these local waveguides have diverse geometries due to the random nature of the fiber, emission spectrum varies as the pump is scanned, exciting different local waveguides across the fiber cross section. We note that we have also analyzed the dependence of the lasing wavelength on the dye material and its concentration, but the results are not presented here because they were consistent with frequently reported characteristic signatures of the dye lasers~\cite{schafer1966organic,vasdekis2007fluidic}. In all results of this section, we have used the same dye material with equal concentration to avoid their impact on our analysis.
\section{Pulse shape}
\begin{figure}[t]
\centerline{
\includegraphics[width=0.85\columnwidth]{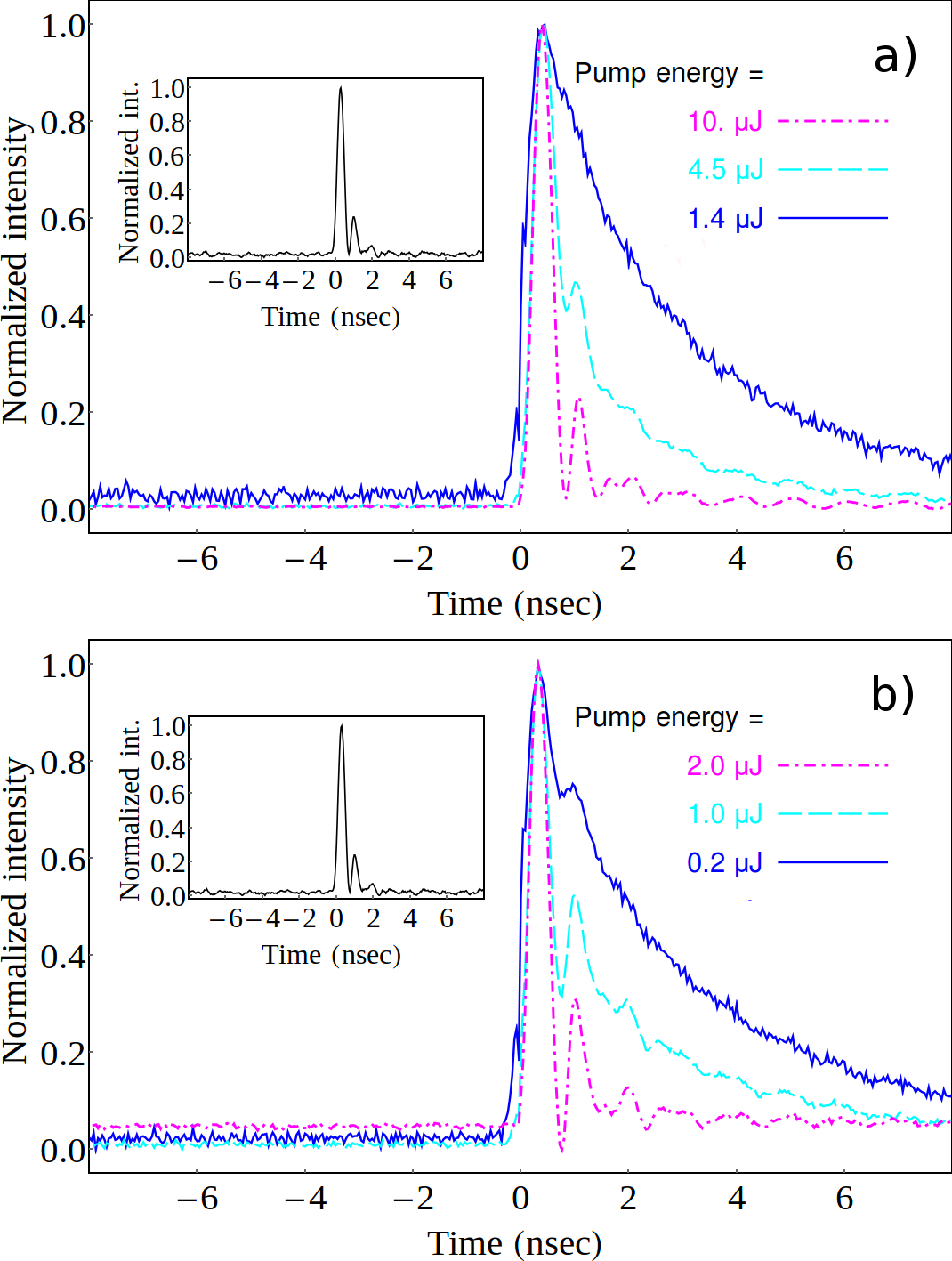}
}
\caption{\label{fig:Time}
{\small \bf Laser pulse shape.} \small Laser pulse in a) Anderson-localized regime, and b) mixed regime of laser operation where the inset is related to the pump pulse. Pulse shortening beyond threshold is apparent in both regimes of laser operation.}
\end{figure}

The laser signal is measured in the time domain using a <300 ps response-time photodetector and an 8 GHz oscilloscope for various pump energies below and above lasing threshold. Fig.~\ref{fig:Time}a shows the laser pulse in Anderson-localized regime and Fig.~\ref{fig:Time}b shows it in the mixed regime. The insets in both figures show the temporal profile of the pump pulse. When the pump pulse temporal front (left-side) hits the dye molecules, the signal rapidly grows as can be seen in the temporal front (left-side) of the signal. Well below the lasing threshold (1.4~\textmu J pump pulse energy in Fig.~\ref{fig:Time}a), the excited dye molecules go through a (relatively slow) spontaneous decay, which explains the slow fall-off in the temporal back (right-side) of the spontaneous emission signal. The exponential decay in the pulse tail has a $\sim 5~ns$ time constant in agreement with the previously reported emission lifetime of rhodamine 640~\cite{gerosa2015all}. In the opposite case when the pump energy is high enough to set the system well above the lasing threshold (10~\textmu J pump pulse energy in Fig. 8a), the laser pulse follows almost exactly the temporal profile of the pump as is expected when the stimulated emission dominates the spontaneous emission in the laser dynamics. The case with 4.5~\textmu J pump pulse energy in Fig.~\ref{fig:Time}a represents the transition region where there is a mixture of stimulated and spontaneous emission. Similar behavior is observed in the mixed regime shown in Fig.~\ref{fig:Time}b. However, the threshold is quite smaller in the mixed regime because of the larger interaction between the gain medium and the optical field due to step-index wave-guiding. The strong interaction results in a larger effective gain hence lowering the lasing threshold.

The reported observations clearly indicate the underlying role of Anderson localization in the lasing behavior of the random fiber laser. The modes of this laser are transversely localized and strongly resemble the modes of the passive system. The dynamics of lasing is dictated by the coupling between the transverse quasi-cavities formed by Anderson-localized modes and the longitudinal Fabry-Perot cavity established by the 4\% reflections at the air-fiber interface of fiber tips. The key observation is directional random lasing mediated by Anderson localization in an optical fiber medium.
Applications of previously reported random lasers are limited mainly because of their multi-directional emission and chaotic fluctuations, where their weak mode confinement results in a high degree of mode competition and therefore chaotic fluctuations. The demonstrated flexible fiber based random laser in this work clearly operates in the regime of Anderson localization leading to a spectrally stable and highly directional random laser. Disorder induced localized states form several isolated channels located across the transverse dimension of g-ALOF. The presence of these guiding channels in conjunction with the Fabry-Perot formed between g-ALOF tips assures a directional random laser. In this implementation, if a narrow pump beam excites one of the disorder induced channels across g-ALOF input facet, the output laser beam follows the transverse position of the pump. On the other hand, under extended input pumping g-ALOF laser results in a laser beam with a very low spatial coherence; we investigate and verify this in the Supplementary by performing the Young double slit experiment~\cite{redding2011spatial,hokr2016narrow}.
 
Recently, random lasers have been applied for speckle free imaging, where it is shown that spatial incoherency helps in avoiding the formation of speckle patterns in optically rough media~\cite{redding2012speckle}. In Ref.~\cite{karbasi2014image}, the randomness in a disordered fiber was used to scramble the incoming spatially coherent laser light and ensured an effective spatial incoherence that improved the image transport metrics. We suggest that using a coherent fiber bundle but with a spatially incoherent laser beam can achieve the same goal, constituting a viable practical application for the system studied in this paper.
The presented work should be viewed as a proof of concept and a proposed platform for potential practical applications. We note that dye-based lasers suffer from optical bleaching; however, it it possible to circulate a dye solution even in a fiber-based platform~\cite{gerosa2015all} or use gases as the gain material~\cite{nampoothiri2012hollow}. A similar effort can be also carried out using Er-doped and Yb-doped random fiber lasers, where the same principals will apply and similar behavior will be expected. 
\section{Conclusions}
The flexible fiber based random laser demonstrated in this paper clearly operates in the regime of Anderson localization leading to a spectrally stable and highly directional random laser. Disorder induced localized states form several isolated channels compactly located across the transverse dimension of g-ALOF. Upon excitation of one of these channels by a narrow input pump, the system starts lasing by the feedback provided through the 4\% reflections at each air-fiber interface. In this implementation, a point to point correspondence between the transverse position of pump and output laser is achieved. Transversely localized laser signal is associated with a more stable frequency response as long as the localization properties are unchanged. The stability of the laser spectrum is attributed to the strong mode confinement provided by the localized states in g-ALOF. 
\section{Conflict of interest}
\noindent
The authors declare no conflict of interest.
\section{AUTHOR CONTRIBUTIONS}
\noindent
B.A. and A.M. wrote the manuscript and all authors contributed to its final editing; J.B. and T.H. fabricated the g-ALOF; B.A. and E.M. conducted all experiments; B.A. analyzed the experimental data, prepared the figures and carried out all the numerical simulations; A.M. conceived the original idea of making an ALOF; B.A., E.M., S.K., and A.M. conceived the idea of using ALOF to make a random laser, and A.M. led the project and supervised all aspects of the work.
\section{Acknowledgments}
\noindent
We are thankful to the UNM Center for Advanced Research Computing, especially Dr. Ryan Johnson, for providing access to computational resources.
\bibliography{refs}{}
\bibliographystyle{naturemag}

\end{document}